\documentclass{article}
\usepackage{amsmath, amssymb, graphicx, hyperref,tabularx,booktabs,ragged2e,natbib}
\usepackage{float}
\usepackage{authblk}

\title{Benchmarking Classical, Machine Learning, and Bayesian Survival Models for Clinical Prediction}
\author[1]{Irving Gómez-Méndez}
\author[1]{Sivakorn Phromsiri}
\author[1]{Ittiphat Kijpaisansak}
\author[1]{Settawut Chaithurdthum}

\affil[1]{CMKL University}
\date{\today}

\begin{document}

\maketitle

\begin{abstract}
Survival analysis is a statistical framework for modeling time-to-event data, particularly valuable in healthcare for predicting outcomes like patient discharge or recurrence. This study implements and compares several survival models—including Weibull, Weibull AFT, Weibull AFT with Gamma Frailty, Cox Proportional Hazards (CoxPH), Random Survival Forest (RSF), and DeepSurv—using a publicly available breast cancer dataset. This study aims to benchmark classical, machine learning, and Bayesian survival models in terms of their predictive performance, interpretability, and suitability for clinical deployment. The models are evaluated using performance metrics such as the Concordance Index (C-index) and the Root Mean Squared Error (RMSE). DeepSurv showed the highest predictive performance, while interpretable models like RSF and Weibull AFT with Gamma Frailty offered competitive results. We also explored the implementation of statistical models from a Bayesian perspective, including frailty models, due to their ability to properly quantify uncertainty. Notably, frailty models are not readily available in standard survival analysis libraries, necessitating custom implementation. Our results demonstrate that interpretable statistical models, when correctly implemented using parameters that are effectively estimated using a Bayesian approach, can perform competitively with modern black-box models. These findings illustrate the trade-offs between model complexity, interpretability, and predictive power, highlighting the potential of Bayesian survival models in clinical decision-making settings. 
\end{abstract}

\section*{Index Terms}
Survival analysis, Bayesian modeling, DeepSurv, hospital length of stay, Cox proportional hazards, random survival forest, Weibull AFT, frailty models.

\section{Introduction}
Statistical survival analysis attempts to predict the amount of time until the occurrence of a particular event, such
as death, recurrence of illness or equipment failure. It is essential in the medical field for assessing treatment effectiveness, comprehending patient prognoses, and allocating healthcare resources as efficiently as possible. Precise survival forecasts allow medical professionals to make well-informed choices, guaranteeing prompt interventions and effective use of medical resources. The primary objective of this study is to evaluate and compare several survival analysis models—classical, machine learning, and Bayesian—for predicting hospital length of stay using censored medical data. We focus not only on predictive performance, but also on the interpretability and uncertainty quantification of each model. Our key contributions include: (1) implementing Bayesian frailty models not readily available in standard toolkits, (2) benchmarking a variety of models using consistent metrics, and (3) discussing the trade-offs between model complexity, interpretability, and clinical applicability.

To guide the reader through this paper, Section 2 introduces the theoretical foundations of survival analysis, including key concepts such as the survival function, hazard function, and likelihood estimation. Section 3 outlines the methodology, describing the dataset, preprocessing steps, and the survival models implemented. Section 4 presents the results, comparing model performance using evaluation metrics like the C-index and RMSE. Section 5 discusses the findings, highlighting the trade-offs between interpretability and predictive power across models. The paper concludes with reflections on practical implications, limitations, and suggestions for future work.

\subsection{Application in Hospital Resource Management}
Reliable survival forecasts are essential for optimizing the distribution of medical resources, including hospital beds, personnel, and equipment, in the context of hospital resource management \cite{kim2024determinants, wen2022time}. Healthcare administrators can better plan interventions, save resources, and enhance patient care by predicting patient outcomes. For example, hospitals can control capacity and optimize operations by forecasting readmission rates or length of hospital stays \cite{vanbelle2021survival}. Utilizing survival analysis, hospitals can:

\begin{itemize}
    \item Allocate ICU beds based on predicted patient deterioration rates,
    \item Optimize discharge planning by forecasting length of stay (LOS),
    \item Identify high-risk patients for early interventions, and
    \item Minimize resource waste by adjusting staff and equipment allocation dynamically.
\end{itemize}

\subsection{Historical Context and Evolution}
Survival analysis has its roots in medical research and actuarial science, with early studies evaluating the effects of treatments and estimating life expectancies. The basis of contemporary survival analysis was established by the 1958 introduction of the Kaplan-Meier estimator, which offered a non-parametric technique for estimating survival functions from censored data \cite{collett2021modelling}. The Kaplan–Meier is, however, a descriptive estimator. It is unable to forecast the survival probability for a new individual based on their specific characteristics (covariates), such as age, tumor size, or treatment type. In other words, it does not incorporate covariate information into its predictions. A semi-parametric approach to evaluating the impact of covariates on survival times without assuming a functional form for the baseline hazard was then provided by the Cox Proportional Hazards (CoxPH) model, which was first presented in 1972 \cite{cox1972regression}. The baseline hazard represents the underlying risk of the event occurring over time for an individual with average or baseline covariates. Not specifying a functional form means that the model does not require this hazard to follow any particular distribution, such as exponential or Weibull, giving CoxPH greater flexibility. Unlike the Kaplan–Meier estimator, the CoxPH model allows for prediction of survival outcomes. As interpretable models that have influenced clinical judgments over the past few decades, these conventional approaches have played a significant role in medical statistics.

As data availability and computational power have increased, survival analysis has expanded to include machine learning-based approaches. Methods like Random Survival Forests (RSF) \cite{ishwaran2008} and DeepSurv \cite{katzman2018} aim to improve predictive accuracy by capturing intricate, non-linear relationships in the data. In contrast, Bayesian approaches represent a separate modeling paradigm. They focus on quantifying prediction uncertainty and incorporating prior knowledge, offering a probabilistic framework that is particularly useful in settings with small sample sizes or complex hierarchical structures. These models have also become more accessible with libraries such as PyMC \cite{salvatier2016pymc3}. For example, traditional parametric models like the Weibull or Weibull AFT can be implemented using a Bayesian approach to estimate full posterior distributions and account for frailty or unobserved heterogeneity.

\subsection{Challenges and Considerations}
Implementing survival analysis in healthcare settings involves several challenges:
\begin{itemize}
\item Data Quality and Censoring: Handling incomplete or censored data—where the event of interest (e.g., death, recurrence, discharge) has not occurred during the observation period— requires robust statistical techniques to avoid biased results.
\item Interpretability: Models must be interpretable to ensure that healthcare professionals can trust and effectively use the predictions in clinical decision-making.
\item Computational Complexity: Advanced machine learning and Bayesian methods require substantial computational resources for model fitting and inference.
\item Ethical and Legal Aspects: The use of patient data necessitates strict adherence to ethical guidelines and data protection regulations to maintain confidentiality and trust.

\end{itemize}

Survival analysis has a vital position in medical studies and hospital administration, and the insights it gives can be used to achieve improved patient outcomes and more efficient healthcare systems. The combination of traditional statistical thinking and modern machine learning and Bayesian approaches is ready to address current challenges and enhance the predictability of survival models. With medical practice increasingly improving, utilization of advanced survival analysis techniques will prove increasingly essential in optimizing resource use and ensuring good patient care.

\subsection{Literature Review}
Existing research in survival analysis has primarily focused on:
\begin{enumerate}
    \item \textbf{Statistical Models} – Cox Proportional Hazards (CoxPH) \cite{cox1972regression}, Weibull regression, and parametric Accelerated Failure Time (AFT) models.
    \item \textbf{Machine Learning Models} – Random Survival Forests (RSF) and Survival Support Vector Machines \cite{vanbelle2021survival}.
    \item \textbf{Deep Learning Models} – DeepSurv and DeepCox \cite{vanbelle2021survival}.
    \item \textbf{Bayesian Models} – Weibull, Weibull AFT, and Weibull AFT with \\Gamma Frailty \cite{collett2021modelling}.
\end{enumerate}

Previous studies have shown that traditional models perform well for small datasets, while machine learning and deep learning models excel in capturing complex patterns. Bayesian models offer robust uncertainty estimation, making them valuable in high-risk applications.
\subsubsection{Time-to-Event Modeling for Hospital LOS Prediction for COVID-19 Patients}

In \cite{wen2022time}, the authors present an in-depth study on hospital length of stay (LOS) prediction using survival analysis techniques. The study emphasizes the critical role of LOS prediction in hospital resource management, patient care planning, and mitigating the strain on healthcare infrastructure, particularly during the COVID-19 pandemic. The authors analyzed 805 COVID-19 patient records from the University of Texas Medical Branch (UTMB), utilizing various survival models to predict individualized LOS distributions. This approach differs from standard regression models by properly handling censored data, ensuring that predictions incorporate patients who were either discharged before reaching an event of interest or remained hospitalized beyond the study period.

The study compared six different time-to-event survival models: Cox Proportional Hazards Model (CoxPH), DeepSurv, Cox-CC (case-control variant of Cox), DeepHit, Multi-task Logistic Regression (MTLR), and Random Survival Forest (RSF). Each model was assessed for its ability to predict hospital LOS using patient demographics, vital signs, and comorbidities. Continuous-time survival models (CoxPH, DeepSurv, and Cox-CC) outperformed discrete-time models (DeepHit, MTLR, and RSF) in terms of predictive accuracy.

To evaluate model performance, they used a range of metrics that reflect both ranking accuracy and prediction error. These include the Concordance Index (C-index), Mean Absolute Error (MAE), Root Mean Squared Error (RMSE), and Mean Absolute Percentage Error (MAPE).
Regarding model performance, RSF achieved the highest C-index score, indicating its effectiveness in ranking patient risk scores accurately. DeepHit and MTLR performed poorly on C-index, likely due to their reliance on discretized time intervals, which introduced bias and instability. When comparing Mean Absolute Error (MAE), Root Mean Squared Error (RMSE), and Mean Absolute Percentage Error (MAPE), Cox-linear and Cox-CC consistently had lower MAE and RMSE values, confirming their robustness for LOS prediction. Deep learning models (DeepSurv, DeepHit) underperformed due to limited training data, as they require large datasets to generalize effectively.

The authors also highlighted several challenges in hospital LOS prediction. First, data scarcity and censorship pose major issues, as many patient records are right-censored—meaning the event of interest (discharge or death) had not yet occurred by the study’s end. Ignoring this can lead to biased predictions. Second, high dimensionality and non-linearity in the data—stemming from complex interactions between patient characteristics, treatments, and disease progression—cannot always be captured by traditional survival models, necessitating more advanced machine learning techniques. Lastly, the skewed LOS distribution further complicates predictions; while most patients are discharged within seven days, some stay significantly longer, making modeling especially difficult for deep learning approaches.

Similar to our project, the authors aimed to develop a predictive tool that hospitals can use for better length-of-stay (LOS) management, reducing bed shortages and improving patient care efficiency. Their paper is particularly relevant to our study, as it underscores the necessity of survival models in handling censored medical data and optimizing hospital resource management. Their findings confirm that traditional survival models (CoxPH and its variants) offer robust and interpretable predictions, which aligns with our goal of creating a clinically useful and interpretable application for hospitals.

\subsubsection{Determinants of Length of Stay for Medical Inpatients Using Survival Analysis}
In the study presented in \cite{kim2024determinants}, the authors investigate the factors influencing hospital length of stay (LOS) among medical patients in Korea. Recognizing the importance of managing LOS for controlling medical costs and optimizing hospital resource utilization, the authors employed survival analysis techniques to identify key determinants affecting patient LOS.

The researchers utilized the National Health Insurance Service (NHIS) sample cohort data from 2016 to 2019, encompassing over 4 million records. They applied the Kaplan–Meier estimator to describe differences in LOS distributions across subgroups defined by characteristics such as sociodemographic factors, patient health status, health checkup results, and institutional attributes. Additionally, the Cox proportional hazards model was used to control for confounding factors, providing a robust analysis of variables influencing LOS.

The study identified several significant determinants of LOS. Older patients exhibited longer LOS, and females had a higher probability of extended hospitalization compared to males. Medical aid recipients experienced longer LOS than those with regional or workplace insurance. Patients in lower income quantiles tended to have prolonged hospital stays. Individuals with severe disabilities had longer LOS compared to those without disabilities. In terms of hospital type, care hospitals were associated with longer LOS, whereas general hospitals showed shorter stays. Furthermore, facilities lacking advanced imaging equipment such as CT, MRI, or PET scanners had higher probabilities of extended patient hospitalization.

Insightful information about the determinants of LOS is provided by this review, including the necessity of having a wide range of determinants, such as organizational factors, patient characteristics, and health status. Time-to-event data for LOS during hospitalization can be legitimately analyzed through the use of survival analysis techniques, such as the Cox proportional hazards model and the Kaplan–Meier estimator. These findings are consistent with our research's objectives, which include improving hospital resource and care planning by accurately estimating LOS.

\section{Theoretical Foundations}

\subsection{Survival, Hazard, and Cumulative Hazard Functions}
In survival analysis, the survival function quantifies the likelihood that a subject will survive beyond a specific time point. It is central to modeling time-to-event data.
It is denoted by the formula:
\[
S(t) = P(T > t),
\]
where \(S(t)\) is the probability of surviving beyond time \(t\), and
\(T\) is the random variable representing the survival time \cite{collett2021modelling}.

The hazard function reflects the instantaneous risk that an event (e.g., death, recurrence) will happen at a specific time, assuming the individual has survived up to that point. It is denoted by the formula:

\[
h(t) = \lim_{\delta t \to 0} \frac{\mathbb{P}(t \leq T < t + \delta t \mid T \geq t)}{\delta t}.
\]

This expresses the conditional probability that the event occurs in the interval \([t, t + \delta t)\), given that it has not occurred before \(t\).  It represents the instantaneous rate at which events occur at time \( t \).

The hazard function can also be expressed in terms of the density and survival functions:

\[
h(t) = \frac{f(t)}{S(t)}.
\]

Here, \( f(t) \) is the probability density function of the event time \( T \). Because f(t) satisfies the relation, \( f(t) = -\frac{d}{dt} S(t) \). An equivalent, and widely used, formulation for the hazard function expresses it as the negative derivative of the log-survival function:

\[
h(t) = \frac{f(t)}{S(t)} = -\frac{d}{dt} \log S(t).
\]

\hspace{\parindent}The cumulative hazard function \(H(t)\) quantifies the total accumulated risk up to time \(t\), and is defined as:

\[
H(t) = \int_0^t h(u) \, du.
\]

The survival function can be derived from the cumulative hazard using:

\[
S(t) = \exp(-H(t)).
\]

These mathematical relationships are foundational in survival analysis and underpin the behavior of models such as CoxPH, Weibull, DeepSurv, and frailty models \cite{collett2021modelling, hazard_function2}.

\subsection{Model Types in Survival Analysis}

Survival models can be broadly classified into three categories: parametric, non-parametric, and semi-parametric. Each category differs in terms of its assumptions about the underlying distribution of survival times and the flexibility it offers.

\begin{itemize}
    \item \textbf{Parametric models} assume that survival times follow a specific statistical distribution, such as the exponential or Weibull distribution. These models are fully specified through distributional parameters and allow for complete likelihood-based inference. They are especially useful when the shape of the hazard is known or well-estimated.
    
    \item \textbf{Non-parametric models}, such as the Kaplan–Meier estimator, make no assumptions about the form of the survival distribution. They are often used for descriptive analysis and provide robust survival estimates, especially when the dataset is large and no prior knowledge about the hazard function shape exists.
    
    \item \textbf{Semi-parametric models}, most notably the Cox Proportional Hazards (CoxPH) model, capture the effect of covariates parametrically but leave the baseline hazard function \( h_0(t) \) unspecified. This provides a balance between flexibility and interpretability and allows for the estimation of relative risk without assuming a particular distribution for survival times.
\end{itemize}

This classification is important because it influences both the interpretability of the model and the types of assumptions that must be made for statistical inference.

\subsection{Censored Data in Medical Research}
One main challenge in survival analysis is handling censored data, which arises when the event of interest has not occurred for some individuals by the end of the study period. In medical studies, censoring occurs because patients can be lost during follow-up, they can drop out from studies, or the studies can be stopped before the event has occurred. Ignoring the censoring observations gives biased outcomes and incorrect inferences. Therefore, there are appropriate statistical methods to correct for censoring and utilize all the information to the fullest \cite{harrell2015}. Mathematically, right-censored data is represented as:

\[
T_i = \min(T_i^*, C_i),
\]
where \(T_i\) is the observed time, \(T_i^*\) is the true survival time, and \(C_i\) is the censoring time.

An indicator variable, \( \delta_i \), is introduced to denote whether an event has occurred:
\[
\delta_i =
\begin{cases}
1, & \text{if the event occurred (uncensored)} \\
0, & \text{if the event is censored}.
\end{cases}
\]

\subsection{Bayesian Approach in Survival Analysis}
A Bayesian approach in survival analysis incorporates prior beliefs or domain knowledge into the modeling process and continuously updates those beliefs using observed data. When used in survival analysis, this approach can enable the model-building process to take into account prior data, professional judgment, or the findings of previous investigations. Bayesian analysis is a natural tool for quantifying uncertainty in parameter estimates and predictions since it expresses model parameters as random variables with known prior distributions $P(\theta)$. In order to accomplish posterior inference, a Bayesian survival model assumes that the parameters are probabilistically distributed:

\[
P(\theta | D) = \frac{P(D | \theta) P(\theta)}{P(D)}
\]

where:
\begin{itemize}
    \item $P(\theta | D)$ is the posterior probability of parameters $\theta$ given the data $D$,
    \item $P(D | \theta)$ is the likelihood — the probability of observing the data $D$ given a specific set of parameters $\theta$,
    \item $P(\theta)$ is the prior probability of $\theta$, and
    \item $P(D)$ is the evidence (marginal likelihood of data).
\end{itemize}

This Bayesian framework is particularly useful in healthcare, where prior domain knowledge can be integrated to improve predictions. Using a Bayesian framework allows creating flexible statistical models that may not be implemented in standard libraries while preserving uncertainty by using the full posterior distribution of the parameters rather than substituting them with single-point estimates \cite{ibrahim2001bayesian}.

\subsection{Partial Likelihood and Log-Likelihood in Survival Analysis}

In the Cox Proportional Hazards (CoxPH) model, we aim to estimate the effect of covariates on the hazard function without specifying the baseline hazard \( h_0(t) \). The hazard for an individual with covariate vector \( \mathbf{X} \) is expressed as:
\[
h(t|\mathbf{X}) = h_0(t) \exp(\boldsymbol{\beta}^T \mathbf{X}),
\]
where \( \boldsymbol{\beta} \) is the vector of regression coefficients. These coefficients are estimated by maximizing the partial likelihood, a method introduced by Cox \cite{cox1972regression} to isolate the covariate effects from the unspecified baseline hazard. The model is considered a form of regression because it expresses the logarithm of the hazard ratio as a linear combination of covariates, with \( \boldsymbol{\beta} \) playing the role of slope parameters, similar to linear or logistic regression.

The partial likelihood relies on the ordering of events and the \textit{risk set} at each failure time. Let \( t_{(1)} < t_{(2)} < \dots < t_{(r)} \) denote the ordered distinct event times, and  \( \mathbf{X}_{(j)} \)  the covariates of the individual who experiences the event at time \( t_{(j)} \). The risk set \( R(t_{(j)}) \) includes all individuals still at risk just before \( t_{(j)} \). The partial likelihood is defined as:
\[
L(\boldsymbol{\beta}) = \prod_{j=1}^{r} \frac{\exp(\boldsymbol{\beta}^T \mathbf{X}_{(j)})}{\sum\limits_{\ell \in R(t_{(j)})} \exp(\boldsymbol{\beta}^T \mathbf{X}_\ell)}.
\]

where:
\begin{itemize}
    \item \( \boldsymbol{\beta} \) is the vector of regression coefficients,
    \item \( \mathbf{X}_{(j)} \) is the covariate vector for the individual who experiences the event at time \( t_{(j)} \),
    \item \( R(t_{(j)}) \) is the risk set just prior to time \( t_{(j)} \), containing all individuals still at risk,
    \item \( \ell \) indexes each individual in the risk set,
    \item \( r \) is the total number of distinct event times.
\end{itemize}

The partial likelihood captures the probability that the individual who actually experienced the event at time \( t_{(j)} \) had the highest relative risk among all individuals still at risk at that time. The numerator represents the risk score (hazard contribution) of this individual, while the denominator sums the risk scores of all individuals in the risk set. By evaluating this ratio at each observed event time and multiplying them together, the partial likelihood isolates the effect of covariates on the hazard, even though the baseline hazard \( h_0(t) \) remains unspecified. This approach allows for consistent estimation of the regression coefficients \( \boldsymbol{\beta} \) based on observed event orderings, without needing to model the exact shape of the underlying hazard function.
Taking the natural logarithm gives the partial log-likelihood:
\[
\ell_p(\boldsymbol{\beta}) = \sum_{j=1}^{r} \left[ \boldsymbol{\beta}^T \mathbf{X}_{(j)} - \log \left( \sum_{\ell \in R(t_{(j)})} \exp(\boldsymbol{\beta}^T \mathbf{X}_\ell) \right) \right].
\]

The partial log-likelihood can be numerically maximized (e.g., via Newton-Raphson) to estimate \( \boldsymbol{\beta} \). Since the baseline hazard function is unnecessary to estimate \( \boldsymbol{\beta} \), this approach allows the Cox model to remain semi-parametric. This is because the baseline hazard function is common to all individuals at risk at a given event time. In conclusion, it does not affect which individual is more likely to fail at that time and thus does not influence the estimation of the covariate effects. As a result, it factors out of the likelihood expression entirely and plays no role in the estimation of \( \boldsymbol{\beta} \) \cite{collett2021modelling}.

In contrast, parametric survival models require explicit specification of the distribution for survival times, such as exponential or Weibull distributions. The parameters of these models, denoted by \( \boldsymbol{\theta} \), are estimated by maximizing the full likelihood function for \( n \) individuals.

\[
L(\boldsymbol{\theta}) = \prod_{i=1}^{n} f(t_i)^{\delta_i} S(t_i)^{1 - \delta_i},
\]
where \( t_i \) is the observed survival time for subject \( i \).

Taking the logarithm yields the log-likelihood:
\[
\ell(\boldsymbol{\theta}) = \sum_{i=1}^{n} \left[ \delta_i \log f(t_i) + (1 - \delta_i) \log S(t_i) \right].
\]

Alternatively, since \( f(t) = h(t) S(t) \), the log-likelihood can also be expressed as:
\[
\ell(\boldsymbol{\theta}) = \sum_{i=1}^{n} \left[ \delta_i \log h(t_i) + \log S(t_i) \right],
\]

This full likelihood is typically maximized using numerical optimization techniques and is used for model comparison, goodness-of-fit testing, and inference through likelihood ratio tests \cite{collett2021modelling}.

These likelihood-based calculations are essential for estimating model parameters and for comparing survival models. The partial likelihood in CoxPH enables consistent estimation of covariate effects without specifying the baseline hazard, making it well-suited for datasets where the form of the hazard function is unknown. In contrast, parametric models rely on the full likelihood, which allows for more direct modeling of survival times when the underlying distribution is known or assumed. Beyond parameter estimation, likelihood-based approaches support broader inference tasks. For instance, full likelihoods are used in likelihood ratio tests, which assess whether adding specific covariates significantly improves model fit. These tools are fundamental for verifying model assumptions and ensuring that survival models remain both accurate and interpretable.

\section{Materials and Methods}

\subsection{Data Description}
The dataset used for this study contains time-to-event data related to recurrence-free survival in breast cancer patients \cite{collett2021modelling}. It includes 686 entries, each representing a unique patient, and contains the following variables:

\begin{itemize}
    \item \textbf{id} – Unique identifier for each patient.
    \item \textbf{treat} – Type of hormonal treatment received (0 = no tamoxifen, 1 = tamoxifen).
    \item \textbf{age} – Age of the patient at the start of treatment (in years).
    \item \textbf{men} – Menopausal status (1 = pre-menopausal, 2 = post-menopausal).
    \item \textbf{size} – Size of the tumor (in millimeters).
    \item \textbf{grade} – Histological grade of the tumor (1 = well differentiated, 2 = moderately differentiated, 3 = poorly differentiated).
    \item \textbf{nodes} – Number of positive lymph nodes.
    \item \textbf{prog} – Progesterone receptor level (in fmol).
    \item \textbf{oest} – Oestrogen receptor level (in fmol).
    \item \textbf{time} – Time to event or censoring (in days).
    \item \textbf{status} – Event indicator (1 = recurrence occurred, 0 = censored).
\end{itemize}

Among the 686 records, 299 are uncensored (event occurred), and 387 are right-censored (no recurrence during the follow-up period). The presence of a substantial proportion of censored observations highlights the importance of using models capable of handling incomplete outcome data.

Preprocessing steps included normalizing continuous variables to ensure numerical stability and encoding categorical features using dummy variables. We then sorted the dataset so that uncensored events appear last and selected the last 75 samples for model testing. The remaining 611 records were used for training.

All models were trained using survival analysis frameworks that natively handle right-censored data, ensuring unbiased parameter estimation. Failure to properly account for censoring can severely impact model calibration and rank-based metrics like the C-index, which we further examine in the discussion.

\subsection{Models Used}

We implemented several survival models, including traditional statistical models, machine learning approaches, and Bayesian approaches. 

In this study, we implemented a mix of existing and custom survival analysis models. Standard models such as Cox Proportional Hazards (CoxPH), Weibull AFT, and DeepSurv were built using existing libraries (Lifelines, PyCox, Scikit-Survival) with minimal modifications. However, the Weibull AFT model with Gamma Frailty was implemented manually using PyMC, as this model is not available in mainstream survival analysis toolkits. We also developed custom evaluation and plotting routines to visualize likelihood contours, survival curves, and model diagnostics in a consistent format across all models. These implementations represent the original technical contributions of this work.

In this study, we define a model as \textit{Bayesian-compatible} if its likelihood function and parameter structure allow it to be naturally expressed in a Bayesian framework using prior distributions and posterior inference. Parametric models such as the Weibull, Weibull AFT, and Weibull AFT with Gamma Frailty fit this criterion because their mathematical formulations include explicit likelihood functions that can be directly combined with prior distributions. These models were implemented using PyMC, enabling full posterior sampling and uncertainty quantification.

In contrast, models like CoxPH, RSF, and DeepSurv are more challenging to implement in a Bayesian framework. CoxPH relies on partial likelihood, which complicates posterior estimation. RSF and DeepSurv, as non-parametric and deep learning-based approaches, lack clearly defined likelihood functions or interpretable priors, making Bayesian implementation non-trivial or impractical. As a result, only the parametric models were fit using Bayesian methods in this study.

\subsection{Implementation Details}

The models were implemented using the following libraries:
\begin{itemize}
    \item \textbf{Lifelines} for traditional survival analysis models (CoxPH, Weibull, Weibull AFT) \cite{lifelines},
    \item \textbf{Scikit-Survival} for Random Survival Forests (RSF) \cite{scikit-survival},
    \item \textbf{PyCox} for deep learning-based survival models like DeepSurv,
    \item \textbf{PyMC} for Weibull AFT Gamma Frailty using a Bayesian approach \cite{salvatier2016pymc3}.
\end{itemize}

\subsubsection{Weibull Model}

The Weibull model is a parametric survival model which assumes that the time-to-event variable follows a Weibull distribution. It is commonly used due to its flexibility in modeling increasing, decreasing, or constant hazard rates over time, depending on the value of its shape parameter.

The hazard function is defined as:
\[
h(t) = \frac{\zeta}{\lambda^\zeta} t^{\zeta - 1}
\]
where:
\begin{itemize}
    \item \( \zeta > 0 \) is the shape parameter — it determines whether the hazard increases (\( \zeta > 1 \)), decreases (\( \zeta < 1 \)), or remains constant (\( \zeta = 1 \)) over time,
    \item \( \lambda > 0 \) is the scale parameter — it stretches or compresses the time axis of the distribution.
\end{itemize}

The survival and cumulative hazard functions are:
\[
S(t) = \exp\left( -\left( \frac{t}{\lambda} \right)^\zeta \right), \quad
H(t) = \left( \frac{t}{\lambda} \right)^\zeta
\]

Given right-censored survival data, the log-likelihood function is:
\[
\ell(\zeta, \lambda) = \sum_{i=1}^{n} \delta_i \log h(t_i) + \log S(t_i)
\]

This model is widely used in clinical and epidemiological research to model time-to-event outcomes \cite{collett2021modelling}.

\subsubsection{Weibull Accelerated Failure Time (AFT) Model}

The Weibull AFT model extends the Weibull model by incorporating covariates that directly influence survival time through a log-linear relationship. The survival time \( T \) is modeled as:

\[
\quad \lambda = \exp\left\{ \beta_0 + \beta_1 X_1 + \cdots + \beta_p X_p \right\}
\]

where:
\begin{itemize}
    \item \(\lambda\) is the location parameter, defined as a linear combination of covariates,
    \item \(\beta_0\) is the intercept term,
    \item \(\beta_1, \dots, \beta_p\) are the regression coefficients,
    \item \(X_1, \dots, X_p\) are the covariate values for an individual,
    \item \(\zeta\) is the scale parameter of the Weibull distribution.
\end{itemize}

This model differs from the Cox Proportional Hazards model in that it explicitly models survival time rather than hazard rates. As a result, the coefficients have a direct interpretation in terms of ``time acceleration or deceleration", making the Weibull AFT model particularly suitable when understanding how covariates stretch or shrink expected survival time is of interest \cite{collett2021modelling}.

\subsubsection{Weibull-Gamma Frailty Model}

The Weibull AFT-Gamma Frailty model is an extension of the Weibull AFT survival model that incorporates unobserved heterogeneity, known as \textit{frailty}, across individuals. The frailty concept refers to unmeasured individual differences that may affect survival but are not captured by observed covariates. These can represent latent risk factors like genetics or lifestyle that increase susceptibility to events. This unmeasured variation is captured using a multiplicative frailty term that accounts for shared or latent risk factors not explicitly included as covariates. In simpler terms, frailty represents hidden influences such as genetic factors, environmental exposures, or lifestyle differences—that can make some individuals more vulnerable to the event, even if they appear similar based on the measured variables.

\medskip

\noindent In this model, the individual hazard function is expressed as:
\[
h(t \mid z) = z \cdot h_{\text{unfrail}}(t),
\]
where \(z\) is the frailty level, and \(h_{\text{unfrail}}(t)\) is the baseline (unfrail) hazard function.

\medskip

\noindent This leads to the following expressions:
\[
H(t \mid z) = z \cdot H_{\text{unfrail}}(t), \quad S(t \mid z) = \exp\{-H(t \mid z)\} = \exp\{-z \cdot H_{\text{unfrail}}(t)\}
\]

Integrating out \(z\), yields the marginal survival function:

\[
S(t) = \int_0^\infty S(t|z) f_Z(z)\,dz.
\]

\medskip

If frailty is modeled as \( z \sim \text{Gamma}(\theta, \theta) \), meaning that \( z \) follows a Gamma distribution with mean 1 and variance \( 1/\theta \), this parameterization ensures the population-level hazard remains centered while allowing for individual variability. Under this assumption, it can be shown that:

\[
S(t) = \left(1 + \frac{1}{\theta} H_{\text{unfrail}}(t) \right)^{-\theta},
\]

\noindent As \(\theta \to \infty\), the variance of the frailty tends to zero, and the model reduces to a standard Weibull model with no unobserved heterogeneity.

\subsubsection{Cox Proportional Hazards (CoxPH) Model}

The Cox Proportional Hazards (CoxPH) model is a cornerstone in survival analysis. It models the relationship between survival time and one or more covariates, while making minimal assumptions about the underlying hazard function. It is a semi-parametric survival model that does not assume a specific distribution for the baseline hazard function. It models the hazard function as:
\[
h(t \mid \mathbf{X}) = h_0(t) \exp\{ \boldsymbol{\beta}^T \mathbf{X} \}
\]

where:
\begin{itemize}
    \item \(h_0(t)\) is the baseline hazard function,
    \item \(\mathbf{X}\) is a vector of observed covariates for an individual,
    \item \(\boldsymbol{\beta}\) represents the regression coefficients associated with covariates \(\mathbf{X}\).
\end{itemize}

\bigskip

%In datasets where multiple individuals experience the event at the same recorded time point (i.e., tied event times), the partial likelihood function must be adjusted. A commonly used approach is Breslow's approximation, which simplifies the computation of the partial likelihood by treating tied events as if they occurred sequentially but without specifying their exact order. This approximation allows efficient estimation of \(\beta\) in the presence of ties and is often used in practice due to its computational simplicity.

The Breslow approximation \cite{breslow1972} estimates the baseline hazard contribution at time \(t_j\) as:

\[
\hat{h}_0(t_j) = \frac{1 - \tilde{\xi}_j}{t_{j+1} - t_j}, \quad \text{with} \quad \tilde{\xi}_j = \exp\left( -\frac{d_j}{\sum\limits_{\ell \in R(t_j)} \exp(\boldsymbol{\beta}^T \mathbf{X}_\ell)} \right)
\]

Where:
\begin{itemize}
    \item \(d_j\) is the number of observed events (deaths) at time \(t_j\),
    \item \(R(t_j)\) is the risk set — the set of individuals still at risk just before \(t_j\).
\end{itemize}

%Breslow’s method simplifies the handling of tied events by assuming they occurred simultaneously and contributes a single term to the partial likelihood denominator. While not as accurate as alternatives like Efron's approximation in the presence of many ties, it remains computationally efficient and widely implemented in survival analysis libraries.

The Cox Proportional Hazards (CoxPH) model is applied in the analysis to assess the effect of covariates on the survival time without assuming a specified baseline hazard distribution. The CoxPH model relies on the proportional hazards assumption, meaning that the hazard ratio between any two individuals is constant over time. In practice, this assumption may not always hold—especially in medical datasets where treatment effects or disease progression can vary over time—but it often provides a reasonable approximation and allows for interpretable estimation of relative risk.

\subsubsection{DeepSurv}

DeepSurv is a neural network-based survival model inspired by the CoxPH framework using the PyCox library \cite{pycox}. It enables the modeling of complex, non-linear interactions between input features while retaining the partial likelihood formulation. It replaces the linear component \(\boldsymbol{\beta}^T \boldsymbol{X}\) with a deep neural network. The hazard function is modeled as:
\[
h(t | X) = h_0(t) \exp\left\{ f_{\theta}(X) \right\},
\]
where \(f_{\theta}(X)\) is the output of a neural network parameterized by \(\theta\), which captures complex, non-linear interactions between covariates.

The model is trained using a loss function derived from the Cox partial likelihood, enabling it to retain the benefits of CoxPH while learning non-linear risk functions. This training process is analogous to maximum likelihood estimation, where the network parameters \(\theta\) are optimized to maximize the partial likelihood. While it is built upon the Cox framework and inherits the proportional hazards assumption, DeepSurv gains greater modeling flexibility by allowing non-linear interactions between covariates.

\subsubsection{Random Survival Forest (RSF)}

Random Survival Forest (RSF) is a non-parametric machine learning approach that extends decision trees to survival analysis \cite{ishwaran2008}. It builds an ensemble of survival trees using bootstrap aggregation (bagging), where each tree is trained on a random sample drawn with replacement from the original dataset. At each internal node, the tree is split based on a statistical test—commonly the logrank test—that evaluates which feature and threshold best separate the survival times of patients into distinct groups. These nodes contain binary decisions, such as “Is tumor size less than 25 mm?”, and the tree recursively splits the dataset to maximize differences in survival between branches. The terminal nodes (leaves) contain survival estimates based on the Kaplan–Meier curves of the samples that fall into that leaf. The logrank test is used at each split to determine which feature split produces the greatest separation in survival distributions.

Each tree in the forest is trained on a different bootstrap sample, and survival predictions are aggregated across trees to improve stability and accuracy. RSF is particularly useful for high-dimensional data where traditional survival models struggle.

\begin{figure}[H]
    \centering
    \includegraphics[width=0.75\textwidth]{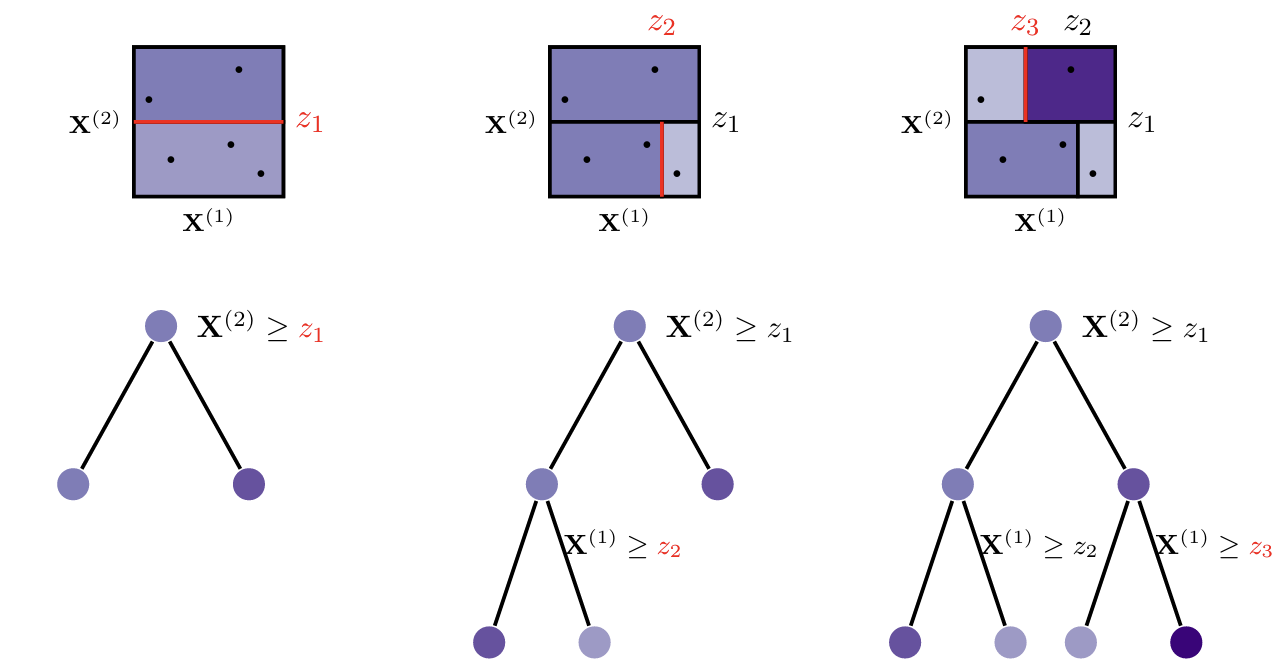}
    \caption{The trees are constructed in a recursive manner. A partition is made over a cell (or, conversely, its corresponding node) at each stage of the tree construction process in order to maximize a split criterion. Adapted from \cite{gomez2021thesis}.}
    \label{fig:RSFSurvival}
\end{figure}

RSF is employed to fit survival data through an ensemble of decision trees to make flexible, data-adaptive survival predictions with no strong parametric assumptions. 
Unlike standard survival models, RSF captures interactions and non-linear effects dynamically and is therefore particularly well-suited to datasets with complex relationships between predictors and survival outcomes. By integrating predictions of a number of survival trees learned on bootstrapped samples, the model achieves prediction stability and avoids overfitting. This is extremely helpful in high-dimensional cases where standard parametric models such as the Weibull or Cox Proportional Hazards model simply cannot learn complex dependencies. The RSF results provide a detailed perspective of survival chances in favor of a data-adaptive approach to survival analysis.

\subsection{Evaluation Metrics}

The performance of each model is assessed using the following metrics:

\begin{itemize}
    \item \textbf{Concordance Index (C-Index)} – Measures the model's ability to correctly rank survival times \cite{brentnall2016cindex}.
    \item \textbf{Root Mean Squared Error (RMSE)} – Quantifies the error in predicted survival times \cite{vanderploeg2014pseudo}.
\end{itemize}

\subsubsection{Concordance Index (C-Index)}

The Concordance Index (C-Index) is one of the most widely used metrics for evaluating survival models. It measures how well a model ranks the survival times of pairs of individuals. The C-Index quantifies the proportion of subject pairs where the model correctly predicts that the individual with a longer observed survival time indeed has a lower risk score \cite{harrell2015}.

Mathematically, the C-Index is computed as:

\[
C = \frac{\sum_{i < j} I(T_i > T_j) I(\hat{T}_i > \hat{T}_j)}{\sum_{i < j} I(T_i > T_j)},
\]

where:
\begin{itemize}
    \item \( T_i, T_j \) are the observed survival times of two individuals \( i \) and \( j \),
    \item \( \hat{T}_i, \hat{T}_j \) are the predicted survival times for individuals \( i \) and \( j \),
    \item \( I(\cdot) \) is the indicator function, which equals 1 if the condition inside is true, and 0 otherwise.
\end{itemize}

A C-Index of 1.0 indicates perfect ranking, meaning the model correctly orders all survival times. A C-Index of 0.5 suggests that the model performs no better than random guessing. Lower values indicate poor predictive performance. 

One limitation of the C-Index is that it does not account for time-dependent prediction performance. It treats all time points equally, potentially underestimating the performance of models that adapt well over time. To address this, time-dependent concordance indices have been developed, which evaluate performance dynamically at different time points rather than as a single global measure.

\subsubsection{Root Mean Squared Error (RMSE)}

Root Mean Squared Error (RMSE) is commonly used to assess the accuracy of survival time predictions. It measures the average magnitude of the difference between the predicted and actual survival times, providing insight into a model’s predictive accuracy \cite{vanbelle2021survival}.

RMSE is defined as:

\[
RMSE = \sqrt{\frac{1}{n} \sum_{i=1}^{n} (T_i - \hat{T}_i)^2},
\]

where:
\begin{itemize}
    \item \( T_i \) is the actual survival time for subject \( i \),
    \item \( \hat{T}_i \) is the predicted survival time,
    \item \( n \) is the number of uncensored observations.
\end{itemize}

A lower RMSE indicates that the model's predictions are closer to the true survival times, reflecting better predictive accuracy. However, RMSE primarily focuses on point predictions and does not account for the probabilistic nature of survival data or censored observations.

When using RMSE in survival analysis, it is essential to properly handle censored data. Since RMSE requires actual survival times, censored observations—where the event has not yet occurred—are typically excluded. This exclusion can lead to biased assessments if censored data are not missing at random \cite{little2019missing}. 

Therefore, while RMSE provides valuable insights into a model’s predictive performance, it should be used alongside other metrics that consider censoring and probability distributions. The combination of C-Index and RMSE offers a comprehensive evaluation of survival models, ensuring that both ranking ability and prediction accuracy are assessed.

\subsubsection{Confidence Intervals for Survival Analysis Predictions}

In survival analysis, estimating confidence intervals for the survivor function \( \hat{S}(t) \) is complicated by the fact that survival probabilities are bounded between 0 and 1. Assuming that \( \hat{S}(t) \) follows a normal distribution is inappropriate, particularly near the tails. A commonly used solution is to apply the complementary log-log transformation:
\[
\log[-\log \hat{S}(t)].
\]

This function maps survival probabilities to the real line, creating an approximately normally distributed variable under regularity conditions.
\break
The standard deviation of the transformed function, \(s_{cl}\), can be estimated using the delta method:
\[
s_{cl} \approx \frac{s}{\hat{S}(t) \cdot |\log \hat{S}(t)|},
\]
where \( s \) is the standard deviation of \( \hat{S}(t) \).

Using this expression, an approximate \( (1 - \alpha) \times 100\% \) confidence interval at a fixed time \(t\) for the transformed value is given by:
\[
\log[-\log \hat{S}(t)] \pm z_{\alpha/2} s_{cl},
\]
where \( z_{\alpha/2} \) is the quantile of the standard normal distribution.

These intervals can be back-transformed to obtain confidence intervals for both the cumulative hazard function and the survival function:
\[
\hat{H}(t) \cdot \exp(\pm z_{\alpha/2} s_{cl}), \quad
\hat{S}(t)^{\exp(\pm z_{\alpha/2} s_{cl})}
\]

This method offers a practical and statistically sound approach to quantify uncertainty in survival predictions at a given time point.

\section{Results}
Table \ref{tab:models} summarizes the models, their loss functions, key parameters, Bayesian compatibility, C-index, and RMSE performance. In this table, the column “Bayesian Fit?” refers to whether the model was estimated using Bayesian inference techniques. A Bayesian fit involves placing prior distributions over the model parameters and using observed data to derive posterior distributions through Bayes' Theorem. Unlike frequentist methods, which yield single-point estimates (e.g., maximum likelihood estimates), Bayesian models generate full probability distributions that capture parameter uncertainty. In this study, models like the Weibull, Weibull AFT, and Weibull AFT with Gamma Frailty were implemented using PyMC to enable Bayesian inference. These fits allow for uncertainty quantification and the integration of prior knowledge, which can be especially valuable in clinical settings with small or heterogeneous datasets.

\bigskip
% \begin{table}[h]
%     \centering
%     \begin{tabularx}{\textwidth}{|X|X|X|c|c|}
%         \hline
%         \textbf{Model} & \textbf{Loss Function} & \textbf{Parameters} & \textbf{Bayesian?} & \textbf{C-Index} \\
%         \hline
%         Weibull & Likelihood & Alpha, Beta & Yes & 0.5 \\
%         Weibull AFT & Likelihood & Alpha, Betas & Yes & 0.612 \\
%         Weibull AFT with Gamma Frailty & Likelihood & Alpha, Betas, Theta & Yes & 0.628 \\
%         CoxPH & Partial Likelihood & Betas & No & 0.594 \\
%         Random Survival Forest & Logrank test & - & No & 0.623 \\
%         DeepSurv & Partial Likelihood & - & No & 0.642 \\
%         \hline
%     \end{tabularx}
%     \caption{Summary of models used}
%     \label{tab:models}
% \end{table}

\begin{table}[h]
    \centering
    \footnotesize % smaller font to make space
    \begin{tabularx}{\textwidth}{>{\raggedright\arraybackslash}X 
                                 >{\raggedright\arraybackslash}X 
                                 >{\raggedright\arraybackslash}X 
                                 c c c} % Added one more 'c' for the new column
        \toprule
        \textbf{Model} & \textbf{Loss Function} & \textbf{Parameters} & \textbf{Bayesian Fit?} & \textbf{C-Index} & \textbf{RMSE} \\ % Example new column
        \midrule
        Weibull & Likelihood & $\zeta$, $\lambda$ & Yes & 0.500 & 1309.04 \\
        Weibull AFT & Likelihood & $\zeta$, $\beta$ & Yes & 0.612 & 1191.63 \\
        Weibull AFT with Gamma Frailty & Likelihood & $\zeta$, $\beta$, $\theta$ & Yes & 0.628 & 1374.13 \\
        CoxPH & Partial Likelihood & $\beta$ & No & 0.594 & 1162.54 \\
        Random Survival Forest & Logrank test & - & No & 0.623 & \textbf{1086.48} \\
        DeepSurv & Partial Likelihood & - & No & \textbf{0.642} & 1158.24 \\
        \bottomrule
    \end{tabularx}
    \caption{Summary of models used. Parameters refer to interpretable coefficients. Models like RSF and DeepSurv do not estimate such parameters explicitly; instead, they rely on learned structures like decision trees or neural networks.}
    \label{tab:models}
\end{table}

\begin{figure}[H]
    \centering
    \includegraphics[width=0.75\textwidth]{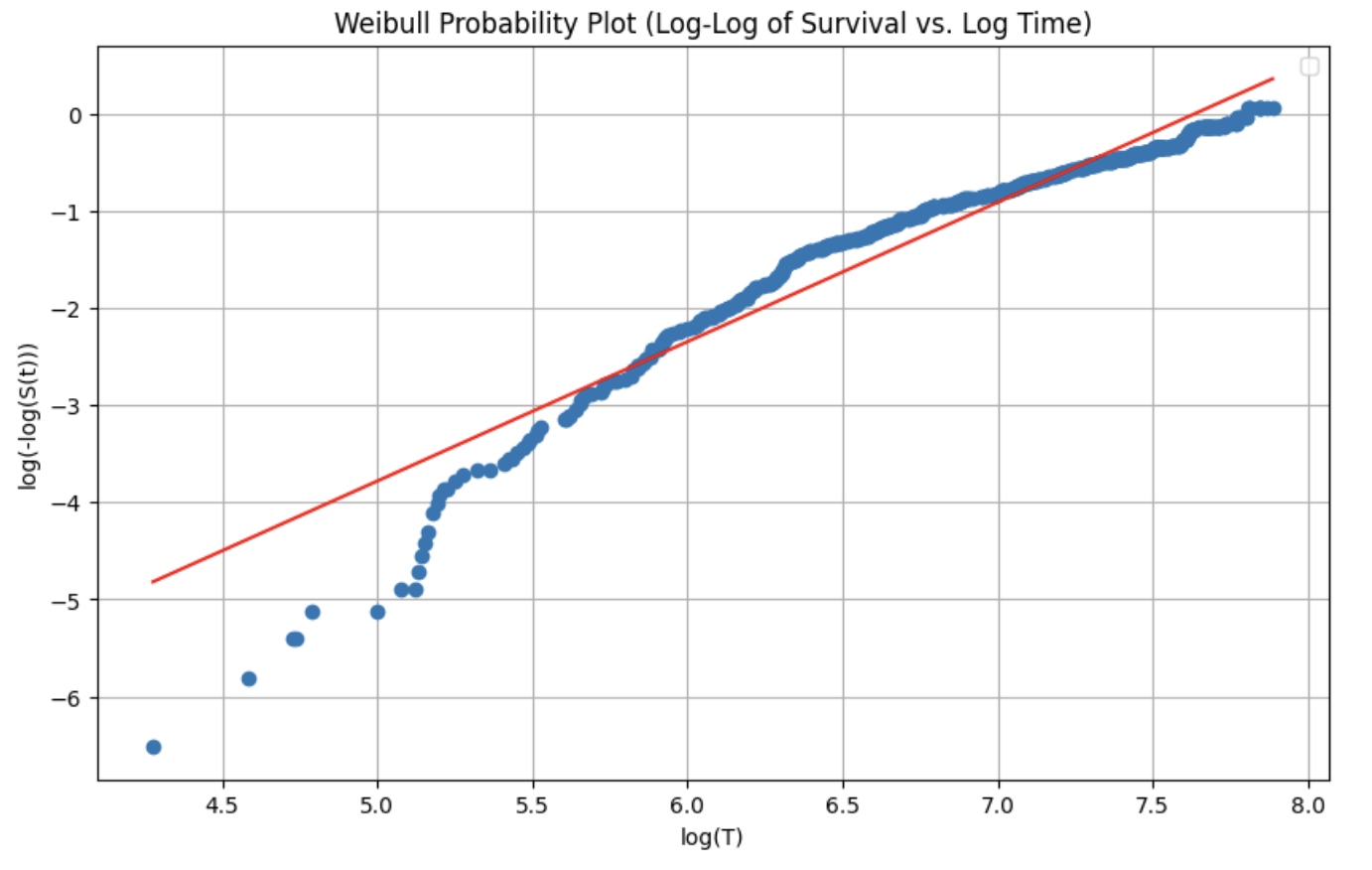}
    \caption{Weibull probability plot with fitted distribution}
    \label{fig:Webb}
\end{figure}

Figure~\ref{fig:Webb} provides graphical validation of the Weibull model fit. Presenting a log-log plot of the empirical survival function versus log-transformed survival time offers a visual check for Weibull suitability. The observed linearity indicates that the Weibull distribution adequately represents the underlying survival process, supporting the appropriateness of using this model in the analysis.

\begin{figure}[H]
    \centering
    \includegraphics[width=0.75\textwidth]{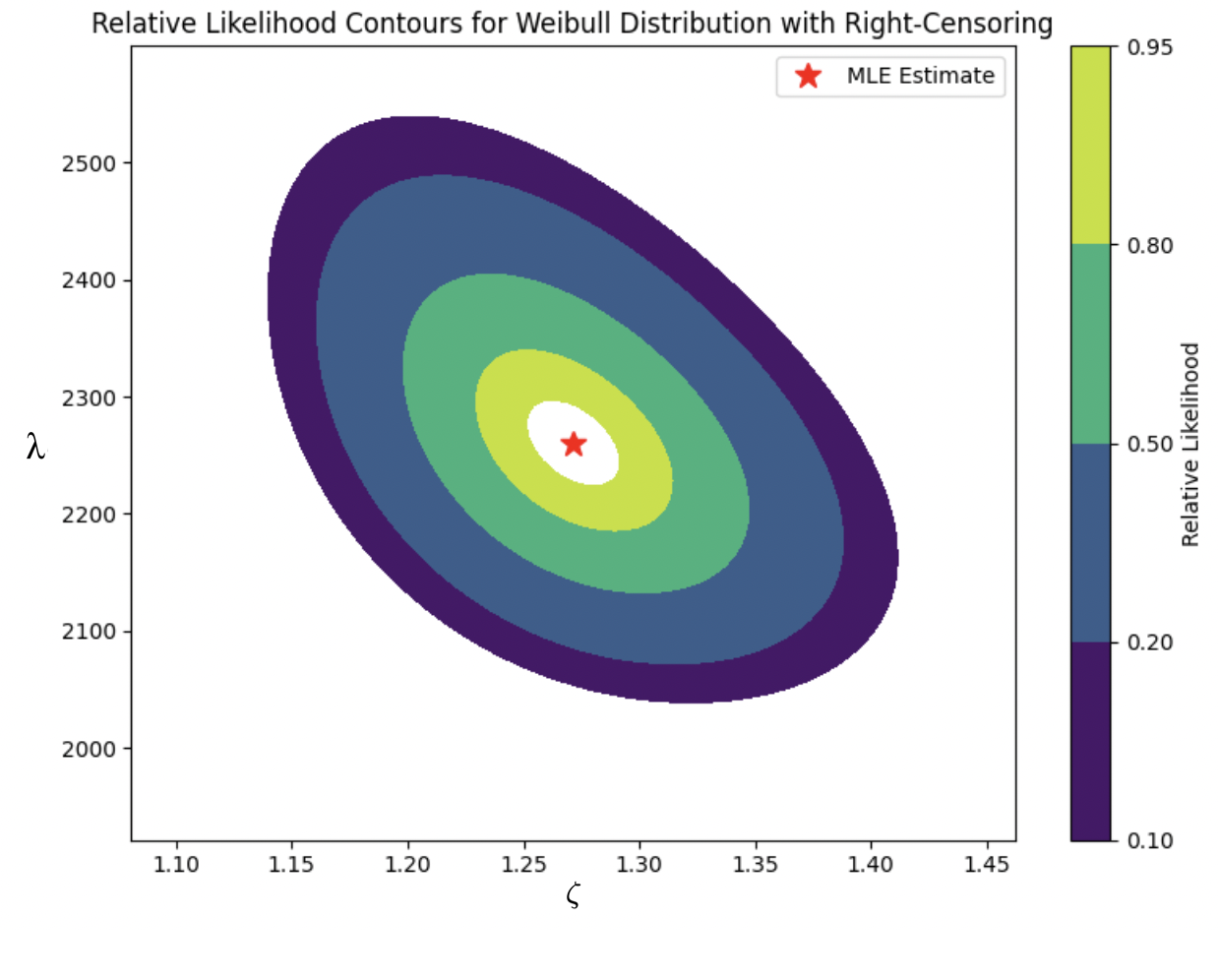}
    \caption{Relative likelihood contours for Weibull distribution}
    \label{fig:Web2}
\end{figure}

Figure~\ref{fig:Web2} displays the relative likelihood contours for the Weibull distribution under right-censoring, where the shape parameter \( \zeta\) is plotted along the x-axis and the scale parameter \( \lambda \) on the y-axis. Each contour represents a region of equal relative likelihood, normalized such that the maximum likelihood estimate (MLE) is the point at which the relative likelihood takes the value 1 (marked by a red star). The color gradient—from dark purple (low likelihood) to yellow-green (high likelihood)—illustrates the concentration of the likelihood around the optimal parameter estimates. These plots help visualize the curvature of the parameter space and illustrate how Bayesian models (especially with Gamma frailty) have more stable and well-defined likelihood peaks, suggesting better convergence and parameter identifiability.

\begin{figure}[H]
    \centering
    \includegraphics[width=0.75\textwidth]{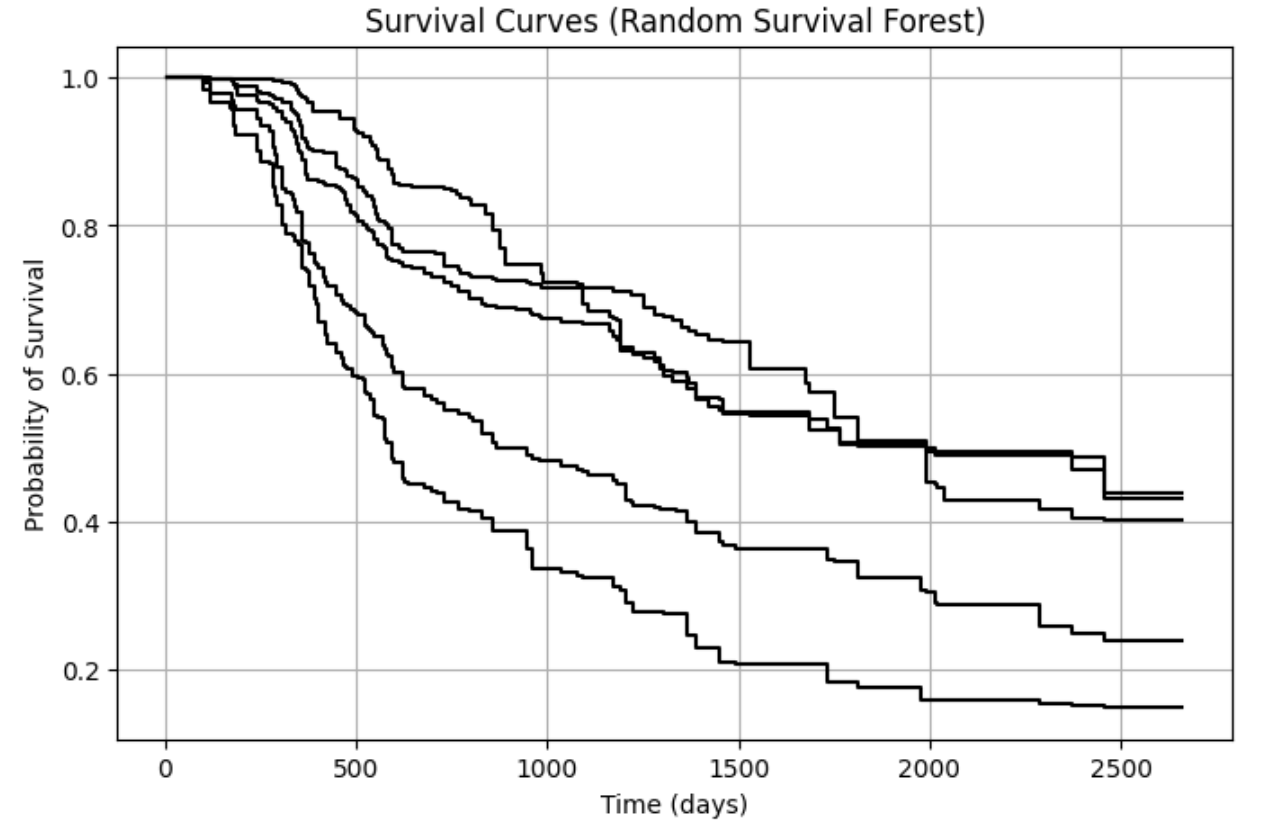}
    \caption{Predicted survival curves using the Random Survival Forest (RSF) model for different patients.}
    \label{fig:RSFSurvival}
\end{figure}

Figure~\ref{fig:RSFSurvival} shows survival functions \(S(t)\) predicted by the Random Survival Forest for a sample of patients. Each curve represents an individual survival estimate, with earlier and steeper drops indicating higher predicted risk. The stepwise form reflects the non-parametric, tree-based estimation used by RSF. Over time, the separation between curves becomes more pronounced, illustrating the model’s ability to stratify patients into different risk profiles.

\begin{figure}[H]
    \centering
    \includegraphics[width=0.75\textwidth]{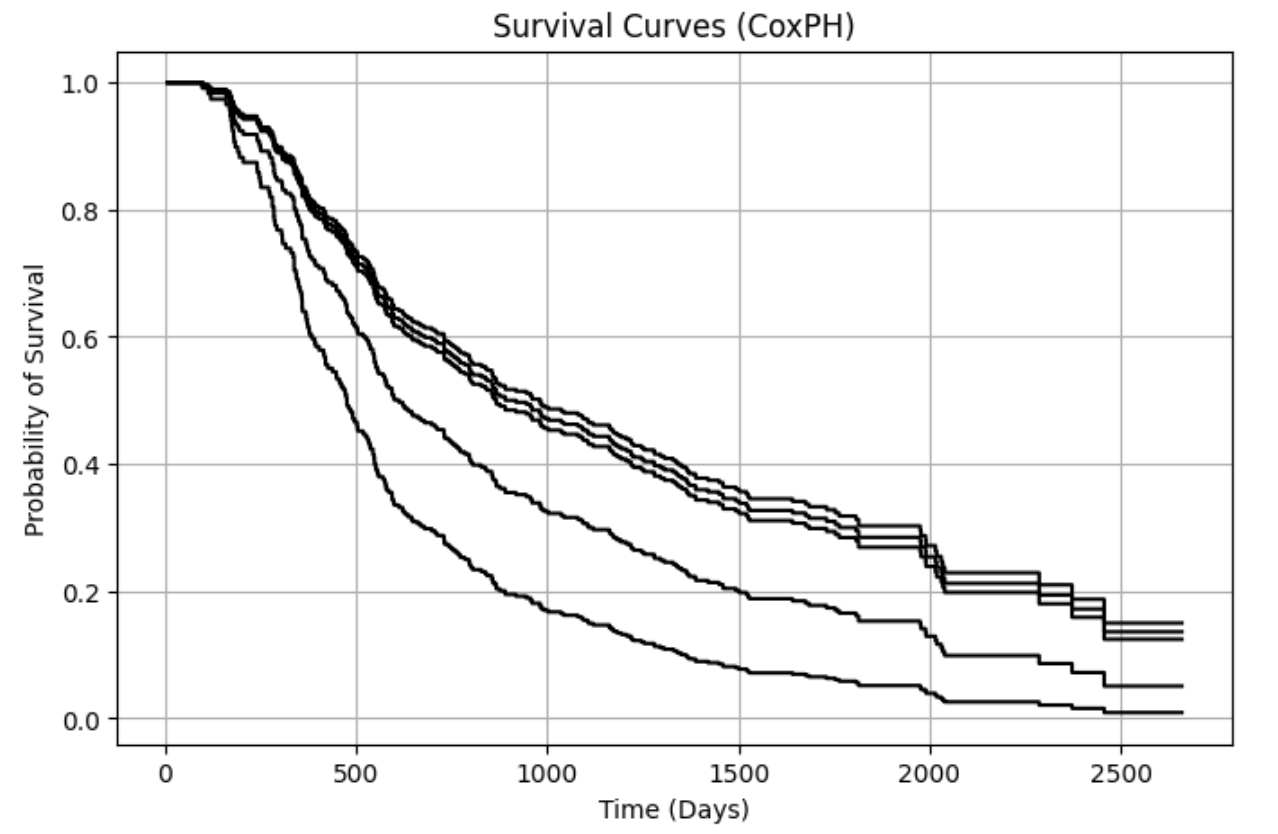}
    \caption{Survival functions for different individuals using the Cox Proportional Hazards (CoxPH) model.}
    \label{fig:Cox}
\end{figure}

Figure~\ref{fig:Cox} shows the survival functions from the CoxPH model, for different individuals. The curves start near 1 and steadily decline over time, showing the probability of survival decreasing as follow-up lengthens. Groups with higher predicted risk drop more quickly, while lower-risk groups decline at a slower pace. The step-like pattern reflects how survival estimates change at observed event times.

\begin{table}[H]
\centering
\begin{tabular}{lc}
\toprule
\textbf{Covariate} & \textbf{Hazard Ratio} \\
\midrule
grade & 1.263 \\
men   & 1.242 \\
nodes & 1.052 \\
size  & 1.009 \\
oest  & 1.000 \\
prog  & 0.998 \\
age   & 0.993 \\
treat & 0.601 \\
\bottomrule
\end{tabular}
\caption{Hazard ratios for covariates based on CoxPH model.}
\label{tab:cox_coefficients}
\end{table}

Table~\ref{tab:cox_coefficients} presents the hazard ratios for each covariate obtained from the Cox Proportional Hazards (CoxPH) model. These values are derived by exponentiating the model’s estimated coefficients, providing a more interpretable representation of effect size. A hazard ratio greater than 1 indicates that the covariate is associated with a higher risk of the event (e.g., recurrence), while a hazard ratio less than 1 suggests a lower risk.

While some models exhibit higher C-index or lower RMSE than others, these differences were not subjected to formal statistical significance testing. As such, performance comparisons should be interpreted with caution. Still, trends across both metrics such as the consistent strength of DeepSurv and the competitiveness of Bayesian Weibull AFT suggest practical differences that may hold in similar datasets. In future work, statistical methods such as bootstrap-based confidence intervals or paired hypothesis tests could be used to rigorously assess whether these performance differences are significant.

While the CoxPH model provides interpretable hazard ratios that highlight the influence of individual covariates, it is equally important to evaluate how different survival models perform in terms of overall predictive accuracy. To follow up, we also compared parametric approaches, focusing on the Weibull AFT model in both its standard and Bayesian variant. This comparison allows us to assess not only interpretability but also the trade-offs in accuracy across different modeling strategies.

\begin{table}[h]
    \centering
    \footnotesize
    \begin{tabular}{lcc}
        \toprule
        \textbf{Model} & \textbf{RMSE} & \textbf{Concordance Index} \\
        \midrule
        Weibull AFT (Bayesian) & 1191.63 & 0.611 \\
        Weibull AFT            & 1308.67 & 0.593 \\
        \bottomrule
    \end{tabular}
    \caption{Performance comparison between the Bayesian Weibull AFT model and the standard Weibull AFT model.}
    \label{fig:WeibullAFT}
\end{table}

Because the RMSE of Weibull AFT with Bayesian method was 1191.63 and the Concordance Index was 0.611, the model was more accurate and enabled better division of high-risk and low-risk patients. It appears that using the Bayesian way in modeling helps by using existing knowledge and accounting for parameter variation, making it a reliable pick for predicting survival outcomes.

\section{Discussion and Conclusion}

Among the evaluated models, DeepSurv achieved the highest C-index of 0.642, demonstrating its effectiveness in capturing complex, non-linear relationships through deep learning (Table \ref{tab:models}). By replacing the linear term of the Cox model with a neural network, DeepSurv gains modeling flexibility at the cost of interpretability, which may pose challenges in domains like healthcare that prioritize explainability. 

Notably, the Weibull AFT model with Gamma frailty achieved a strong second place with a C-index of 0.628 (Table \ref{tab:models}). It ended up outperforming some machine learning techniques such as RSF while retaining full interpretability, making it particularly valuable in clinical applications (Figures 2 and 3 show the survival curves and likelihood surfaces for this model).

Random Survival Forest (RSF) followed closely with a C-index of 0.623 (Table \ref{tab:models}). As a non-parametric ensemble method, RSF effectively handles high-dimensional data and captures feature interactions without relying on parametric assumptions. Its predictive power makes it a strong candidate for general-purpose use, although its ``black-box" nature may limit interpretability and user trust in clinical environments.

Beyond overall performance, models like RSF and CoxPH also offer insights into feature importance. In our dataset, the number of positive lymph nodes, tumor size, and hormone receptor levels (progesterone and oestrogen) emerged as consistently influential predictors across multiple models. In particular, a higher node count and larger tumor size were associated with increased hazard, while higher hormone receptor levels tended to correlate with longer recurrence-free survival. These patterns align with clinical understanding of breast cancer progression.

With a C-index of 0.612, the Weibull AFT model showed a good balance between interpretability and predictive performance. It is particularly helpful in situations where comprehending the impact of predictors is just as crucial as producing precise forecasts because it directly models survival time through covariates. The semi-parametric and well-known CoxPH model produced a marginally lower C-index of 0.594. Despite this, because of its theoretical foundation, interpretability, and broad use in biostatistics, it continues to be a fundamental tool in survival analysis. 

With a C-index of 0.500, the conventional Weibull model—which ignores covariates—performed the worst. This finding emphasizes the limited usefulness of simple models in intricate, patient-specific datasets and the significance of incorporating covariates to improve prediction accuracy.

Interestingly, when presented in a Bayesian context, all three of the Bayesian-compatible models—the standard Weibull, Weibull AFT, and Weibull AFT with Gamma Frailty—performed better than their frequentist counterparts (Table \ref{fig:WeibullAFT}). Specifically, the Bayesian Weibull AFT with Gamma Frailty outperformed its non-frailty counterpart, achieving a C-index of 0.628. These models were able to better handle small sample sizes and unobserved heterogeneity by incorporating priors and posterior uncertainty, especially in clinical settings where patient variability is high. Although Bayesian models might be more computationally demanding, they are extremely useful for reliable, comprehensible survival analysis because of their capacity to express uncertainty and take into account past knowledge. 

Censoring plays a critical role in determining model reliability and the accuracy of predicted survival functions. Models that explicitly incorporate censored observations—such as CoxPH, RSF, and our Bayesian implementations—tend to produce more calibrated survival curves and hazard estimates. Conversely, models that underutilize censored data or rely heavily on complete cases may exhibit biased risk rankings and systematically underestimate or overestimate survival time, especially in the tail regions of the distribution.

Despite promising results, several practical limitations must be considered. Deep learning models like DeepSurv require large datasets to generalize well, which can be a barrier in many clinical settings with limited sample sizes. Bayesian models, while powerful, demand greater computational resources and expertise, making them harder to deploy without appropriate infrastructure. Additionally, many survival models assume that censoring is non-informative—a condition that may not always hold true in hospital settings, where early discharge or dropout may be systematically linked to patient condition. These factors should be weighed when selecting models for real-world deployment.

In terms of feature importance, models like CoxPH provide interpretable outputs that reveal the influence of individual covariates on survival outcomes. As shown in Table~\ref{tab:cox_coefficients}, the number of positive lymph nodes is associated with increased hazard (hazard ratio = 1.052), consistent with clinical expectations. Tumor grade and menopausal status also indicate elevated risk, with hazard ratios of 1.263 and 1.242, respectively. The treatment variable shows a substantial protective effect, with a hazard ratio of 0.601. In contrast, hormone receptor levels and age exhibit hazard ratios close to 1, suggesting limited impact on recurrence risk. These results demonstrate that CoxPH can reveal clinically meaningful patterns and align with known prognostic factors in breast cancer.

The results indicate variations in performance among the models. Bayesian approaches offer robust uncertainty quantification while retaining transparency, whereas traditional models remain easy to understand. On the other hand, deep learning models exhibit strong predictive potential but require large datasets. Notably, the Weibull AFT with Gamma Frailty model was implemented manually using PyMC, as no standard library provided this capability. Its performance demonstrates that custom Bayesian modeling can compete with well-established machine learning methods.

Overall, the findings show that interpretability and model complexity are traded off (Table~\ref{tab:models} compares predictive performance, while Figures 2–5 illustrate model fit, survival patterns, and parameter uncertainty. Although DeepSurv and RSF exhibit high predictive accuracy, their adoption in fields that demand transparency may be constrained by their difficulty in explanation. On the other hand, Bayesian and statistical models like Weibull AFT and CoxPH provide interpretable frameworks that are still helpful in clinical decision-making. These results demonstrate that interpretable statistical models, when properly implemented and estimated using Bayesian methods, can compete with or even outperform more complex machine learning techniques. This highlights the importance of selecting survival models that not only perform well but also align with the practical and ethical needs of the application domain. In summary, while several models were implemented using existing libraries, our custom development of the Weibull-Gamma Frailty model and related visualizations demonstrate the feasibility and impact of original survival modeling efforts in constrained medical settings.

\section*{Note}

Codes to reproduce our results are available in
\href{https://github.com/R-Ohm/ANN-Hospital-LOS}{https://github.com/R-Ohm/ANN-Hospital-LOS}

\bibliographystyle{unsrt}
\bibliography{references}

\end{document}